\documentclass[]{pasj02}

\Received{}
\Accepted{}
 
\usepackage{xcolor}
\usepackage[switch,mathlines]{lineno}
\usepackage{amsmath}

\begin{document} 

\title{Evidence for Charge Exchange Emission in Supernova Remnant N132D from XRISM/Resolve Observations

}


\author{Liyi Gu\altaffilmark{1,2,3,4}\orcid{0000-0001-9911-7038}\email{l.gu@sron.nl}}
\author{Hiroya Yamaguchi\altaffilmark{5}}  
\author{Adam Foster\altaffilmark{6}}
\author{Satoru Katsuda\altaffilmark{7}}
\author{Hiroyuki Uchida\altaffilmark{8}}
\author{Makoto Sawada\altaffilmark{9}}
\author{Frederick Scott Porter\altaffilmark{10}}
\author{Brian J.\ Williams\altaffilmark{10}}
\author{Robert Petre\altaffilmark{10}}
\author{Aya Bamba\altaffilmark{11}}
\author{Yukikatsu Terada\altaffilmark{7,5}}
\author{Manan Agarwal\altaffilmark{12}}
\author{Anne Decourchelle\altaffilmark{13}}
\author{Matteo Guainazzi\altaffilmark{14}}
\author{Richard Kelley\altaffilmark{10}}
\author{Caroline Kilbourne\altaffilmark{10}}
\author{Michael Loewenstein\altaffilmark{15,10,16}}
\author{Hironori Matsumoto\altaffilmark{17}}
\author{Eric D.\ Miller\altaffilmark{18}}
\author{Yuken Ohshiro\altaffilmark{11,5}}
\author{Paul Plucinsky\altaffilmark{6}}
\author{Hiromasa Suzuki\altaffilmark{5}}
\author{Makoto Tashiro\altaffilmark{7,5}}
\author{Jacco Vink\altaffilmark{12}}
\author{Yuichiro Ezoe\altaffilmark{19}}
\author{Ehud Behar\altaffilmark{20}}
\author{Randall Smith\altaffilmark{6}}

\altaffiltext{1}{SRON Space Research Organisation Netherlands, Niels Bohrweg 4, 2333 CA Leiden, the Netherlands}
\altaffiltext{2}{Leiden Observatory, Leiden University, PO Box 9513, 2300 RA Leiden, The Netherlands}
\altaffiltext{3}{RIKEN High Energy Astrophysics Laboratory, 2-1 Hirosawa, Wako, Saitama 351-0198, Japan}
\altaffiltext{4}{Department of Physics, Tokyo University of Science, 1-3 Kagurazaka, Shinjuku-ku, Tokyo 162-8601, Japan}
\altaffiltext{5}{Institute of Space and Astronautical Science (ISAS), Japan Aerospace Exploration Agency (JAXA), Kanagawa 252-5210, Japan}
\altaffiltext{6}{Center for Astrophysics -- Harvard-Smithsonian, MA 02138, USA}
\altaffiltext{7}{Department of Physics, Saitama University, Saitama 338-8570, Japan}
\altaffiltext{8}{Department of Physics, Kyoto University, Kyoto 606-8502, Japan}
\altaffiltext{9}{Department of Physics, Rikkyo University, Tokyo 171-8501, Japan} 
\altaffiltext{10}{NASA / Goddard Space Flight Center, Greenbelt, MD 20771, USA} 
\altaffiltext{11}{Department of Physics, University of Tokyo, Tokyo 113-0033, Japan} 
\altaffiltext{12}{Anton Pannekoek Institute, the University of Amsterdam, Postbus 942491090 GE Amsterdam, The Netherlands}
\altaffiltext{13}{Université Paris-Saclay, Université Paris Cité, CEA, CNRS, AIM, 91191, Gif-sur-Yvette, France}
\altaffiltext{14}{European Space Agency (ESA), European Space Research and Technology Centre (ESTEC), 2200 AG, Noordwijk, The Netherlands} 
\altaffiltext{15}{Department of Astronomy, University of Maryland, College Park, MD 20742, USA} 
\altaffiltext{16}{Center for Research and Exploration in Space Science and Technology, NASA / GSFC (CRESST II), Greenbelt, MD 20771, USA} 
\altaffiltext{17}{Department of Earth and Space Science, Osaka University, Osaka 560-0043, Japan} 
\altaffiltext{18}{Kavli Institute for Astrophysics and Space Research, Massachusetts Institute of Technology, MA 02139, USA} 
\altaffiltext{19}{Department of Physics, Tokyo Metropolitan University, Tokyo 192-0397, Japan} 
\altaffiltext{20}{Department of Physics, Technion, Technion City, Haifa 3200003, Israel}




\maketitle

\begin{abstract}
XRISM has delivered one of its first light observations on N132D, the X-ray brightest supernova remnant in the Large Magellanic Cloud. Utilizing 193~ks of high-resolution X-ray spectroscopy data, we conduct a comprehensive search for charge exchange emission. By incorporating a charge exchange model into our spectral analysis, we observe an improvement in the fits of two weak features at 2.41~keV and 2.63~keV. These features, with a combined significance of 99.6\%, are consistent with transitions from highly ionized silicon ions in high Rydberg states, which are unique indicators of charge exchange. Our analysis constrains the charge exchange flux to no more than 4\% of the total source flux within the 1.7–3.0 keV band, and places an upper limit on the charge exchange interaction velocity at 450~km s$^{-1}$. This result supports ongoing shock-cloud interactions within N132D and highlights the unique capabilities of XRISM to probe the complex physical processes at play.


\end{abstract}


\section{Introduction}

Astrophysical charge exchange typically involves the transfer of one or more electrons from a donor atom to a highly ionized ion, such as O$^{7+}$ and Fe$^{26+}$. This process generates diagnostic X-ray emission, including an enhanced forbidden line of the He$\alpha$ triplet and cascade lines from high principal quantum number ($n$) states (see \cite{gu2023} for a recent review). Charge exchange X-ray emission has been observed in the near-Earth environment, including interactions between the solar wind and various celestial bodies such as planets and comets \citep{lisse96, cravens2000, br2007, dennerl2010}. Beyond interactions with the solar wind, evidence suggests that charge exchange might occur in more distant regions, including novae \citep{mitrani2024}, supernova remnants (SNRs, \cite{katsuda2011,uchida2019}), starburst galaxies \citep{liu2011}, and clusters of galaxies \citep{gu2015,gu2017}. While these observations are still tentative, the detection of charge exchange phenomena holds significant potential for studying the interplay between cold and hot matter at different scales.

The advent of the non-dispersive high-resolution spectrometer Resolve \citep{ishisaki2022} onboard the X-Ray Imaging and Spectroscopy Mission (XRISM, \cite{tashiro2024}) offers a groundbreaking opportunity to uncover the presence of charge exchange in objects beyond the solar system. Resolve allows us to perform spatially resolved measurements of the forbidden-to-resonance ratios and to probe enhanced cascade lines from high $n$. The high-$n$ transitions often serve as a more robust fingerprint of charge exchange, whereas alternative explanations exist for the forbidden-to-resonance ratio \citep{gu2016b,hitomirs2018}. To further search for evidence of charge exchange, we can examine spatial correlations between these lines and the neutral clouds that often serve as electron donors.

The first light observation using Resolve, N132D, presents an ideal opportunity for searching for evidence of charge exchange. N132D is the X-ray brightest SNR in the Large Magellanic Cloud (LMC), at a distance of 50~kpc \citep{pietrzyski2019}. Evidence suggesting the presence of charge exchange in N132D comes from infrared and radio studies, which reveal its interaction with dense molecular clouds \citep{w2006,sano2015,dopita2018,sano2020}. While recent investigations with the XMM-Newton Reflection Grating Spectrometer (RGS) have suggested the presence of charge exchange based on the O and Ne triplet line ratios \citep{suzuki2020}, the RGS is unable to isolate the high Rydberg lines due to line blending and instrumental broadening. The Resolve instrument offers a brand new window for the search for charge exchange, by fully resolving the transitions of Si, S, Ar, Ca, and Fe up to high Rydberg states \citep{xrism2024}. 

This paper is organized as follows. In \S\ref{sec2}, we present the first light observation, the spectral analysis and the results. Our findings are discussed and summarized in \S\ref{sec3} and \S\ref{sec4}. Throughout the paper, the errors are given at a 68\% confidence level.

\section{Observation, Analysis, and Results}
\label{sec2}

XRISM was launched on September 7th, 2023 from JAXA's Tanegashima Space Center. N132D was observed twice, on 3 December and 9 December, for a total Resolve duration of 204~ks (OBSID = 000128000 and 000126000, PROCVER = 03.00.011.008, TLM2FITS = 004\_001.15Oct2023\_Build7.011). Since the Resolve aperture door (``gate valve'') has not opened yet, a 250-$\mu$m-thickness beryllium filter attenuates X-ray photons in the soft X-ray band, reducing the effective area \citep{midooka2020}. This restricted the bandpass to energies above 1.7~keV.

The data reduction follows the procedure outlined in \citet{xrism2024}, the XRISM collaboration paper on N132D thermal structure (hereafter "the thermal paper"), with only a concise summary provided here. The Resolve detector gain was calibrated on-orbit using 72 measurements of $^{55}$Fe sources during Earth occultation, achieving an energy resolution of 4.43~eV (FWHM) and a scale error of 0.04~eV at 5.9 keV. The calibration pixel, which is continuously illuminated by $^{55}$Fe, maintained an FWHM of 4.39~eV and an energy scale error of 0.11~eV throughout the observation period under the same calibration scheme. This confirms the performance of the instrument during the main observation, even outside the gain fiducial intervals. Systematic uncertainties on the energy scale were $\sim 0.3$~eV in the 5.4–8.0~keV band and up to 1.0~eV between 1.7~keV and 5.4~keV \citep{porter2024}. After the screening, the total exposure time was slightly reduced to 193~ks. Only "high-resolution primary" events are used for the spectral analysis. The non-X-ray background has been modeled, while the sky background is ignored due to its negligible contribution to spectral region above 2~keV. To account for instrumental effects, a redistribution matrix file was created using the {\texttt rslmkrmf} task with the cleaned event file and CALDB derived from ground-based measurements. The line-spread function model incorporated a Gaussian core, low-energy exponential tails, escape peaks, and Si fluorescence lines. An auxiliary response file was generated using the {\texttt xaarfgen} task, assuming a point-like source at the aim point as input. The spectral files remain identical to those used in the thermal paper. Spectral fitting was performed using SPEX software version 3.08.01 \citep{kaastra1996} along with an updated atomic database.


The Resolve spectrum has been optimally binned using the approach proposed in \citet{kb16}. The optimal bin size varies depending on factors such as spectral resolution, the number of resolution bins, and the local intensity of the spectrum, resulting in different bin sizes for each energy range. Typically, we employ a bin size of 3~eV for energies below 7~keV, while for higher energies, where the flux decreases, a bin size of 4.5~eV is utilized. Although the bin size is getting close to the spectral resolution of the detector, the spectral features are still well sampled in practice due to the significant Doppler line broadening present in this object. 

To explore potential weak features in the Resolve spectrum of N132D, we begin by analyzing the full array spectrum for a global picture of its thermal structure. 

\subsection{Thermal structure}

The thermal structure in N132D has been explored previously using a combination of multiple collisionally ionized components \citep{hughes1998, behar2001, borkowski2007}. Using 240~ks of Suzaku data combined with a 60~ks NuSTAR observation, \citet{bamba2018} successfully reproduced the broadband spectrum with a model comprising two thermal components, with a 0.7 keV component in ionization equilibrium and a 1.5 keV component in an over-ionized state, plus a power-law component. By analyzing a 200~ks XMM-Newton RGS spectrum, \citet{suzuki2020} characterized the thermal structure using a three-component non-equilibrium ionization (NEI) model. Their analysis revealed temperatures of 0.2 keV, 0.56 keV, and 1.36 keV, with an ionization timescale of approximately $10^{11}$ cm$^{-3}$ s.

In the thermal paper, the temperature structure of the SNR has been analyzed using the same XRISM/Resolve data, complemented by the imaging capabilities of the Xtend data. The preferred model, referred to as `model B', comprises three components:  a cool $\sim$0.8~keV ionizing component with modest velocity broadening $\sim 450$ km s$^{-1}$ for the soft X-ray lines, an intermediate $\sim$1.8~keV ionizing component with large velocity dispersion $\sim 1670$ km s$^{-1}$ for most of the Fe He$\alpha$ lines, and a hot 10~keV component with broadening $\sim 750$ km s$^{-1}$ for the Fe Ly$\alpha$ lines. The main driver for the three component model is that
it is challenging for a simpler model to reproduce both
the ratio of the Fe Ly$\alpha$ to Fe He$\alpha$ lines and the width of Fe Ly$\alpha$ lines that appear to be narrower than the He$\alpha$ lines. We adopt the same thermal model for this work. The abundances of Si, S, Ar, Ca, and Fe are tied among the three components.

The thermal structure modeling is performed using three NEI components in SPEX, providing an opportunity to crosscheck the results obtained in the thermal paper with the AtomDB model. The best-fit parameters for the cool component are a temperature of $0.79 \pm 0.01$~keV, an ionization timescale of $0.9 \pm 0.5 \times 10^{12}$ cm$^{-3}$ s, a radial velocity of $230 \pm 40$ km s$^{-1}$, and a velocity dispersion of $450 \pm 30$ km s$^{-1}$.

For the intermediate component, the best-fit temperature is $1.9 \pm 0.1$~keV, with a lower limit on ionization timescale of $ 1.0 \times 10^{12}$ cm$^{-3}$ s. The velocity dispersion is $1650 \pm 150$ km s$^{-1}$, and the radial velocity is $320 \pm 80$ km s$^{-1}$. The properties of the hot component are fixed to the best-fit values derived in the thermal paper.

These best-fit parameters are consistent within uncertainties with the AtomDB-based values reported in the thermal paper, demonstrating generally good agreement between the SPEX and AtomDB models on this object. The global fit yields a C-statistic value of 2196, matching the expected C-statistic of 2239 within its root mean square deviation of 67. The C-statistic/dof of 2196/2091 shows a marginal improvement over the value of 2430/2232 reported in the thermal paper. This difference likely comes from subtle differences in the optimal binning routines and atomic data used in SPEX and AtomDB. A detailed assessment of the effect from atomic data will be presented in a followup paper.

\subsection{Search for charge exchange features}

\begin{figure*}
  \begin{center}
\includegraphics[width=1\textwidth]{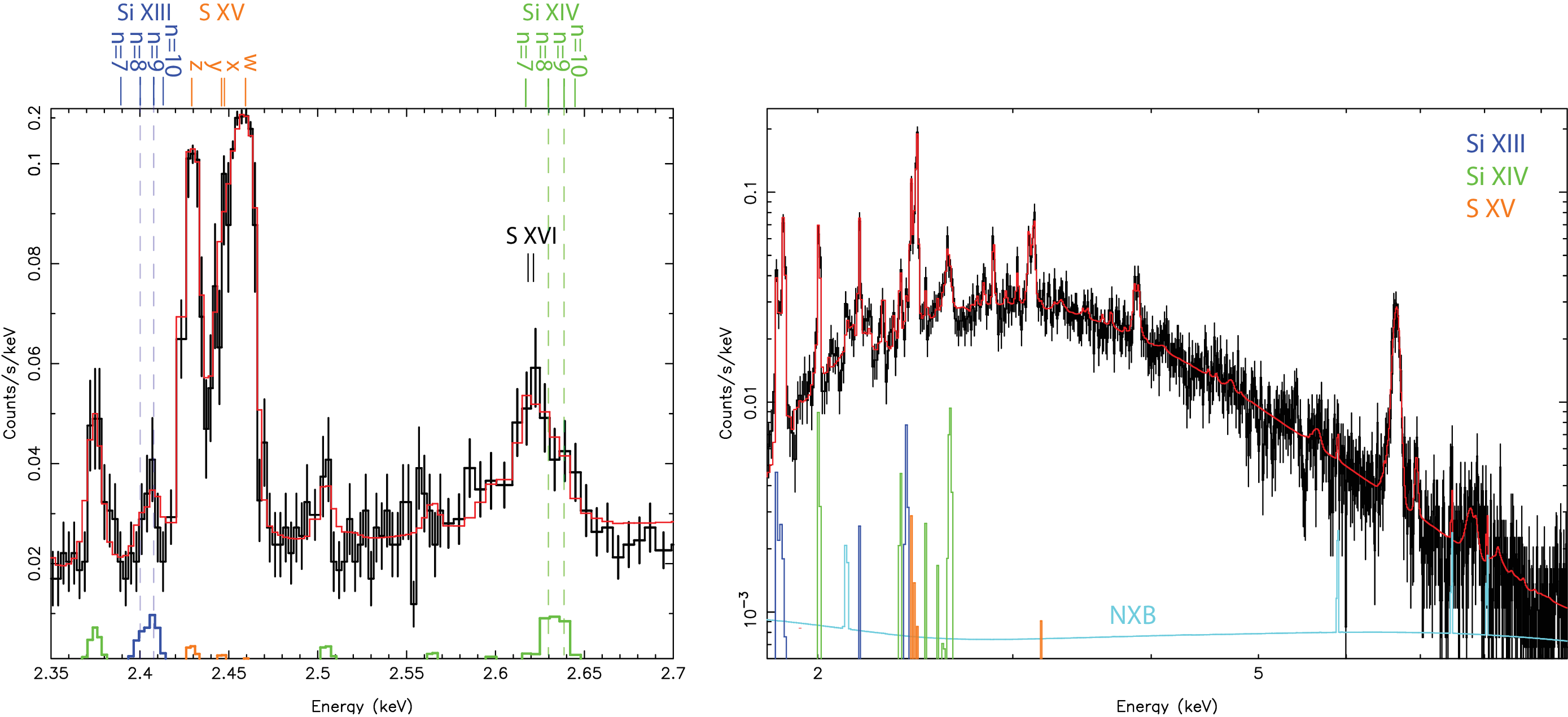}
  \end{center}
  \caption{Best-fit model for the XRISM/Resolve spectrum of N132D, including three {\tt nei} components and one {\tt cx} component. (Left): The charge exchange emission from Si~XIII and Si~XIV, shown in blue and green, respectively, reproduces the weak features at 2.41~keV and 2.63~keV. (Right): Broad-band view of the charge exchange emission in N132D, highlighting the Si~XIII, Si~XIV, and S~XV lines in blue, green, and orange, respectively. The non-X-ray background is shown in cyan. Alt text: Two graphs show energy levels in keV with various spectral lines labeled for Si XIII, Si XIV, Si XV, and S XV. The left graph focuses on Si and S, and the right graph provides a broader spectrum. }
  \label{fig:cxplot}
\end{figure*}

The above baseline model incorporating three thermal components describes well the strong line features in the Resolve spectrum. However, there may still be weak features that require further modeling. To explore the potential presence of charge exchange emission, we incorporate a {\tt cx} component \citep{gu2016a} into the fit, allowing the emission measure and collision velocity free to vary. The temperature, velocity dispersion, and abundances for the {\tt cx} component are fixed to those of the cool thermal component.

The atomic cross-section data for the H- and He-like charge exchange emissions are sourced from the multi-channel Landau–Zener calculations within the Kronos version 3.1 database \citep{mullen2016}. The He-like data are fully resolved up to levels indicated by orbital angular momentum $l$, while the H-like data are only resolved by the principal quantum number $n$ due to the significant degeneracy of the $l$ states inherent in the Landau–Zener framework. To enable line calculation, we assume an ad hoc $l$-distribution. Common choices include the ``statistical'', ``separable'' \citep{smith2014}, and ``low-energy'' \citep{krasnopolsky2004} distributions. The ``statistical'' distribution significantly underestimates the strengths of the 8$p$ $\rightarrow$ 1$s$ and 9$p$ $\rightarrow$ 1$s$ transitions, while the ``separable'' and ``low-energy'' distributions provide better agreement with observations and yield similar spectra for N132D. We therefore adopt the ``low-energy'' distribution for this analysis.

Inclusion of the {\tt cx} component improves the fit, reducing the C-statistic to 2181 for the expected value of 2239, with a $\Delta$C of 15 relative to the baseline model. The most significant improvement occurs at 2.41~keV, where a weak emission feature is well accounted for by charge exchange emission between H-like Si and atomic hydrogen, resulting in He-like Si transitions from $n=8$ and $n=9$. Another improvement is observed at 2.63~keV, potentially corresponding to $n=8$ and $n=9$ transitions from H-like Si, arising from interactions between bare Si and atomic hydrogen. These states with high principal quantum numbers are difficult to populate via electron-impact processes, making it unlikely that these lines could be produced by a thermal component.

The significance of incorporating the {\tt cx} component into the spectral model can be evaluated using the Akaike Information Criterion (AIC, \cite{aic1974}). As shown above, the inclusion of the {\tt cx} component yields a $\Delta$C improvement of 15 with the addition of two parameters, corresponding to a significance level of 99.6\%. The total charge exchange flux has an upper limit of $5.7 \times 10^{-13}$ erg cm$^{-2}$ s$^{-1}$ in the $1.7-3.0$~keV band, accounting for 4\% of the total source flux at the 68\% confidence level. In addition, the fit provides a marginal constraint on the collision velocity of the charge exchange component, setting an upper limit of 350 km s$^{-1}$. Higher collision velocities would suppress the lines from $n=8$ and $n=9$ in the model, resulting in a poorer fit to the observed data. The line-of-sight velocity of the charge exchange lines is consistent within uncertainties with that of the cool thermal component.

The high ionization state of the possible {\tt cx} component makes it unlikely to be spectral contamination from solar wind charge exchange, as the ionization temperature of solar wind is typically ten times lower \citep{kuntz2019}, and the flat light curve of the N132D observation indicates an absence of contamination from solar activity.  

These features are unlikely to originate from radiative recombination continua, as the expected positions for recombination edges are at 2.44~keV for Si~XIII and 2.67~keV for Si~XIV. Explaining the observed features as recombination would require a redshift of $\sim 0.01$, which is inconsistent with that measured in the other detected Si transitions.

The inclusion of the possible {\tt cx} component influences the measurement of the thermal structure, in particular the metal abundances. When the {\tt cx} component is added to the baseline model, the inferred abundances of Si and S decrease by 20\% and 4\%, respectively, compared with those obtained using the baseline model. These differences exceed the statistical uncertainties of 6\% for Si and 3\% for S, highlighting the importance of accounting for charge exchange when measuring abundances in the thermal plasma. Although the {\tt cx} component calculates all elements, the current model predicts negligible contributions from charge exchange for ions beyond Si XIII, Si XIV, and S XV, as illustrated in Fig.~\ref{fig:cxplot}. Consequently, the inclusion of the {\tt cx} component primarily affects the abundances of Si and S, with a negligible impact on other elements.

Given its ionization temperature of 0.8~keV, the charge exchange component observed by Resolve contributes even less to soft X-ray lines, with only $\sim2$\% to the O~VIII and Ne~X lines. This value aligns roughly with the results of \citet{suzuki2020} using XMM-Newton RGS. The absence of Fe~XXV and Fe~XXVI charge exchange lines suggests that the interaction of the hot components with neutrals is likely less significant. 

We have conducted a parallel analysis using the XSPEC package, incorporating the non-equilibrium ionization models and charge exchange model {\tt ACX2} provided by AtomDB \citep{foster2012, foster2020}. The emission excesses at 2.41 keV and 2.63 keV are successfully modeled by Si$^{12+}$ and Si$^{13+}$ transitions within {\tt ACX2}, yielding a similar improvement in the C-statistic. Furthermore, the Si$^{12+}$ and Si$^{13+}$ line ratios are consistent with an interaction between the cool thermal component and neutral matter, at an impact velocity less than $450$ km s$^{-1}$. These results are consistent with those obtained using SPEX-CX.

To robustly confirm charge exchange signatures from Si lines in N132D, an additional 360~ks XRISM observation with the gate valve closed, or 40~ks if the gate valve opens, would be required to achieve a 5$\sigma$ confidence level.

\bigskip

\section{Discussion}
\label{sec3}

Using the first light Resolve observation of N132D, we conducted a search for potential charge exchange emission features. Enhanced high Rydberg transitions of Si~XIII and Si~XIV are observed at 2.41~keV and 2.63~keV, with a combined statistical significance of 99.6\% compared with a thermal-only model. While these transitions are weak, they represent signatures of charge exchange resulting from electron capture into the n=8 and n=9 states. These lines offer a much more direct indication of charge exchange than previously reported forbidden-to-resonance line ratios in n=2 transitions. These findings highlight the capability of Resolve to identify subtle features, even with the gate valve closed, and suggest its potential for future studies of charge exchange in astrophysical environments.

We further investigate the physical properties, location, and scale of regions emitting charge exchange, where hot gas interacts with neutrals. Based on the best-fit emission measure for the {\tt cx} component, the interaction volume $V$ could be estimated by
\begin{equation}
\label{cxeq}
V \approx 5\times 10^{45}  \left(\frac{10^{3} \; {\rm cm^{-3}}}{n_{\rm N}}\right) \left(\frac{1 \; {\rm cm^{-3}}}{n_{\rm I}}\right)  {\rm m^{3}},
\end{equation}
where $n_{\rm N}$ and $n_{\rm I}$ are the hydrogen densities of neutral and ionized clouds. One candidate location is the CO-emitting clouds reported by \citet{sano2020} with high CO $J$ = 3–2 / 1–0 intensity ratios ($>$1.5), suggesting shock-cloud interaction. Assuming neutral and ionized hydrogen densities of 700 cm$^{-3}$ and 5 cm$^{-3}$, the estimated interaction volume is approximately $5 \times 10^{-5}$ pc$^3$, about $2 \times 10^{-6}$ of the total CO cloud volume.

Alternatively, the charge exchange emission could originate from the boundary layer between the ionized gas shell and the surrounding atomic hydrogen cloud. For a shell diameter of approximately $\sim$25~pc and density of $\sim$30~cm$^{-3}$ \citep{sano2020}, Eq.~\ref{cxeq} yields an interaction volume of roughly $1\times 10^{-3}$~pc$^3$ and a minimal interface thickness of $5\times 10^{-7}$~pc. This thickness is clearly smaller than the mean free path of atomic hydrogen in the ionized cloud ($\sim 10^{-3} - 10^{-5}$~pc). 

A third potential source is the shell-like dust structure reported by \citet{rho2023}. In this scenario, swept-up dust grains from the blast wave could contribute to charge exchange emission through direct interaction between ions and dusts \citep{kharchenko2012}, or through ion interactions with secondary atoms produced by shock evaporation \citep{itoh1989}.

Previous high-resolution X-ray spectroscopy using the XMM-Newton RGS has revealed anomalously high forbidden-to-resonance ($f/r$) line ratios as a common feature in SNRs. However, the limited angular resolution and sensitivity above 2 keV of the RGS have prevented definitive conclusions about their origin. Proposed mechanisms include resonance scattering \citep{vanderHeyden2003, amano2020}, charge exchange \citep{katsuda2012, tanaka2022, koshiba2022}, recombining plasmas \citep{vanderHeyden2003, broersen2011}, and low-temperature plasma \citep{tanaka2022, koshiba2022}. A joint analysis using Resolve and RGS data offers a promising path to uncovering the processes behind these $f/r$ anomalies.

\section{Summary}
\label{sec4}

Using the first light 193~ks XRISM/Resolve observation of the LMC SNR N132D, we searched for charge exchange emission. Our analysis indicates a potential charge exchange interaction between the N132D shock and surrounding neutral clouds, with a moderate collision velocity of $\leq 450$ km s$^{-1}$. This interaction appears to produce high Rydberg transitions in Si XIII and Si XIV, with a combined significance of 99.6\%. These results offer valuable insights into shock-cloud interactions and demonstrate the potential of XRISM/Resolve to detect charge exchange lines and other spectral features in a wide range of astrophysical environments.


\begin{ack}
SRON is supported financially by NWO, the
Netherlands Organization for Scientific Research. This work was supported by JSPS Core-to-Core Program (grant number:JPJSCCA20220002).
\end{ack}


\end{document}